# The Waterwheel in the Waterfall

D. Balciunas

**Abstract**. A fundamental problem in evolutionary ecology research is to explain how different species coexist in natural ecosystems. This question is directly related with species trophic competition. However, competition theory, based on the classical logistic Lotka-Volterra equations, leads to erroneous conclusions about species coexistence. The reason for this is incorrectly interpreted interspecific interactions, expressed in the form of the competition coefficients. Here I use the logistic Lotka-Volterra type competition equations derived from the so called resource competition models to obtain the necessary conditions for species coexistence. These models show that only species with identical competitive abilities may coexist. Due to such relations between competing species ecosystems biodiversity decreases in the course of evolution.

World ecosystems are rich in species. But how is species diversity maintained? I reduce this problem to the following question. When do competing species coexist? Considering the formal description of species trophic competition we encounter with two kinds of interactions. Either populations of $n$ species $x_i$ use a common resource or $m$ species of resources $R_j$ are exploited by the same consumer. Traditionally only the first case biologists call competition. The second one was named apparent competition (Holt, 1977). I use the terms divergent competition (d-competition) and convergent competition (c-competition) for them respectively. $c^n d^m$-competition describes a fully connected food web of $n$ consumers and $m$ resources.

Suppose for simplicity that the total mass density $M = \Sigma R_j + \Sigma x_i$ of the above food web components remains constant. Then c- and d-competitive interactions may be formalized by a set of equations

$$dR_j/dt = \rho R_j - \Sigma \beta_i^{\,j} x_i \qquad (1)$$

$$dx_i/dt = \beta_i x_i - q_i x_i$$

where $\rho$, $\beta_i^{\,j}$, $\beta_i$ and $q_i$ are functions such that $\Sigma \rho R_j = \Sigma q_i x_i$ and $\beta_i = \Sigma \beta_i^{\,j}$. Functions $\beta_i^{\,j}$ (and $\beta_i$) describe mass transformation due to interactions between $R_j$ and $x_i$. From (1) I derive the relative competition strength functions $\Phi_k$ for both types of interactions

$$\Phi_k = \mathbf{c}\mathbf{u} + \mathbf{d}_k\mathbf{u}, \qquad k = 1, ..., p, \qquad p \in \{m, n\} \qquad (2a)$$

or

$$\Phi_k = \Phi_{ind} + \Phi_{dir,k} \qquad (2b)$$

Here $\Phi_k$ is equal to $(\varphi_k u_k - du_k/dt)/\varphi_k u_k$; $\varphi_k$ is some function of the growth rate of population mass density. The terms on the right side of the equation (2a) represent the relative strength of the two sides of competition process. I call them indifferent and directed competition, respectively. $\mathbf{u}$ is a vector of either consumers $x_i$ or resources $R_j$ densities expressed in the same units. $\mathbf{c}$ and $\mathbf{d}_k$ are vectors with the components $c_k$ and $d_{kl}$ ($l = 1, ..., k$). $c_k$, an indicator of indifferent competition, describes the competitive ability of a population density unit of $k$th species. $d_{kl} = c_k - c_l$, an indicator of directed competition, may be interpreted as a competition potential. $c_k$ has the following expressions

$$c_j^{-1} = R_j \Sigma q_i x_i / \Sigma \beta_i^{\,j} x_i \qquad \text{for c-competitors} \qquad (3)$$

$$c_i^{-1} = M - q_i \Sigma R_j / \beta_i \qquad \text{for d-competitors}$$

Equations (2) have the same form as the logistic Lotka-Volterra competition models, employed as a mathematical description of Darwinian evolutionary theory (Gause, 1934). However, as they were not related with resources exploitation explicitly (Grower, 1997), their interpretation was incorrect and wrong conclusions have been drawn using these classical competition models. Particularly the errors are related with the description of the nature of interspecific competition coefficients, usually indicated as $\alpha_{kl}$. As we see from (3) all such coefficients in the model (2) are equal to 1. As a result of this misunderstanding the theory of limiting similarity arose (MacArthur and Levins, 1967).

When the system (2) reaches a stationary state where the densities of all populations are finite and stable, $\Phi_k = 1$ and $\Phi_{ind} = 1$ by the definition. Then $\Phi_{dir,k} = 0$ for all $k$. Thus, species do not exclude each other from a model community if only $\mathbf{Du} = \mathbf{0}$, where $\mathbf{D}$ is a skew-symmetric matrix with the components $d_{kl}$. The solution of this matrix equation, satisfying $\mathbf{u} > \mathbf{0}$, is $\mathbf{d}_k = \mathbf{0}$ ($k = 1, ..., p$). This result means that only identical species, which have the same competitive abilities $c_k$, coexist. A community of identical species is in the evolutionary steady state where only indifferent competitive interactions occur. It is worth to note here that although the possibility of the stable coexistence of competing identical species is mentioned from time to time in various contexts (Richerson *et al.*, 1970; Holt, 1977; Ågren and Fagerström, 1984; Abrams, 1986; Walter, 1988; Cornell and Lawton, 1992; Tilman and Pacala, 1993), it was never realized that this way of coexistence, as the expression of a general principle of nature, was the only possible.

If $c_k$'s are not constants then the densities of coexisting species fluctuate. In the case of some cyclic trajectory W we have the following relations, which describe the evolution of the densities of all populations:

$$w^{-1} \int_W \Phi_k \, dw = 1, \qquad k = 1, …, p \tag{4}$$

$$w^{-1} \int_W \Phi_{ind} \, dw = 1 \tag{5}$$

$$w^{-1} \int_W \Phi_{dir,k} \, dw = 0, \qquad k = 1, …, p \tag{6}$$

and finally

$$w^{-1} \int_W d_{kl} \, dw = 0, \qquad k = 1, …, p \tag{7}$$

A system of competing species evolves to the state where the mean differences between competing species (7) and the mean pressure of directed competition (6) disappear. Depending on a particular model we may obtain more or less complicated dynamics of coexisting species (Huisman and Weissing, 1999; Kuang *et al.*, 2003).

In ecosystem with constant total mass the changes of the overall biomass of coexisting species reflect alterations in species competing abilities. If a particular community retains its biomass stability, it indicates that species in this community are identical. This will add to the discussions on ecosystems biodiversity-stability issues (Hughes and Roughgarden, 2000).

Theoretically the unlimited number of c- and d-competitors may coexist. One of the factors that prevent them to increase their number is the ability of species to reach the identity. Fortunately, species involved in $c^n d^m$-competition have not to be identical with respect to every resource or consumer to coexist. It is enough that they would be identical regarding the whole number of resources or consumers. What is more, if every consumer species do not exploit its own unique combinations of resources, some $\beta_i^j = 0$. Then $c_k$'s depend on the different sets of parameters (3). This enhances the probability of coexistence and leads to the growth of the number of competing species. Consequently it makes a space for the increasing of species richness within the opposite trophic level.

Only competition potentials drive Darwinian evolution of species. Indifferent competition has no direct effects on evolutionary events. Although Darwin understood that individual variations were a necessary prerequisite for selection to occur, he did not realize that the differences between individuals competitive abilities itself were a force of evolution. Instead of this he erroneously proposed that the unlimited reproduction of living beings induced selection due to the 'struggle for life' (Darwin, 1859). This took him away from the idea about what is really going on in ecosystems. Indeed, Darwin too much centred on the subject of artificial selection and his theory is very much an artificial selection theory. Even if the results of evolution in nature and of artificial selection look like the same, there is a considerable difference between them. Founded on the ground of artificial selection, Darwin's evolution theory is like the waterwheel in the waterfall creating an order by sifting the best adapted species and destroying less favourable variants. Here the 'waterfall' and 'waterwheel' are the symbols of the 'struggle for life' and Darwin's natural selection. To take natural selection as a steam engine is not a strange point of view in science (Løvtrup, 1976). Meanwhile, because ecosystems move toward such an equilibrium where all $\mathbf{d}_k = \mathbf{0}$, Darwinian evolution is merely the waterfall between the levels set by the value of competition potential $d_{kl}$. Such an evolutionary force as natural selection really does not exist in nature.

The affirmation that the biodiversity of the living world is huge is true only from our viewpoint. Ecosystems would think quite the reverse. The only criterion on which they can judge is the flow of energy along food chains. The chains may converge and diverge if only the competing routes retain the same mass-transforming value expressed as $c_k$. The evolution of ecological communities goes in such a direction that diverse species concerning their energetic relations with the heterogeneous environment form a homogeneous system.

**References**

Abrams, P. A. (1986). The competitive exclusion principle: other views and a reply. *Trends Ecol. Evol.* **1**, 131-132.

Ågren, G. I. and T. Fagerström (1984). Limiting dissimilarity in plants: randomness prevents exclusion of species with similar competitive abilities. *Oikos* **43**, 369-375.

Cornell, H. V. and J. H. Lawton (1992). Species interactions, local and regional processes, and limits to the richness of ecological communities: a theoretical perspective. *J. Anim. Ecol.* **61**, 1-12.

Darwin, C. (1859). *On the Origin of Species*, London: John Murray.

Gause, G. F. (1934). Experimental analysis of Vito Volterra's mathematical theory of the struggle for existence. *Science* **79**, 16-17.

Grower, J. P. (1997). *Resource Competition*, London: Chapman & Hall.

Holt, R. D. (1977). Predation, apparent competition, and the structure of prey communities. *Theor. Popul. Biol.* **12**, 197-229.

Hughes, J. B. and J. Roughgarden (2000). Species diversity and biomass stability. *Am. Nat.* **155**, 618-627.

Huisman, J. and F. J. Weissing (1999). Biodiversity of plankton by species oscillations and chaos. *Nature* **402**, 407-410.

Kuang, Y., W. F. Fagan and I. Loladze (2003). Biodiversity, habitat area, resource growth rate and interference competition. *Bull. Math. Biol.* **65**, 497-518.

Løvtrup, S. (1976). On the falsifiability of neo-Darwinism. *Evol. Theory* **1**, 267-283.

Richerson, P., R. Armstrong and C.R. Goldman (1970). Contemporaneous disequilibrium, a new hypothesis to explain the "paradox of the plankton". *Proc. Natl. Acad. Sci.*USA **67**, 1710-1714.

Tilman D. and S. Pacala (1993). The maintenance of species richness in plant communities, in *Species Diversity in Ecological Communities*, R. E. Ricklefs and D. Schluter (Eds), Chicago: University of Chicago Press, pp. 13-25.

Walter, G. H. (1988). Competitive exclusion, coexistence and community structure. *Acta Biotheor.* **37**, 281-313.